\documentclass[12pt]{JHEP}

\usepackage{epsfig}

\newcommand{\newc}{\newcommand}
\newc{\R}{$R$}
\newc{\charginom}{M_{\tilde \chi}^{+}}
\newc{\mue}{\mu_{\tilde{e}_{iL}}}
\newc{\mud}{\mu_{\tilde{d}_{jL}}}
\newc{\barr}{\begin{eqnarray}}
\newc{\earr}{\end{eqnarray}}
\newc{\beq}{\begin{equation}}
\newc{\eeq}{\end{equation}}
\newc{\ra}{\rightarrow}
\newc{\lam}{\lambda}
\newc{\eps}{\epsilon}
\newc{\gev}{\,GeV}
\newc{\tev}{\,TeV}
\newc{\eq}[1]{(\ref{eq:#1})}
\newc{\eqs}[2]{(\ref{eq:#1},\ref{eq:#2})}
\newc{\etal}{{\it et al.}\ }
\newc{\eg}{{\it e.g.}\ }
\newc{\ie}{{\it i.e.}\ }
\newc{\nonum}{\nonumber}
\newc{\lab}[1]{\label{eq:#1}}
\newc{\dpr}[2]{({#1}\cdot{#2})}
\newc{\gsim}{\stackrel{>}{\sim}}
\newc{\lsim}{\stackrel{<}{\sim}}
\newc{\ol}{\overline}
\newc{\thr}{\theta}
\newc{\mq}{m_{\tilde{Q}_3}}
\newc{\md}{m_{\tilde{D_3}}}
\newc{\thq}{{\theta_{\tilde{Q_3}}}}
\newc{\thd}{{\theta_{\tilde{D_3}}}}
\newc{\thg}{{\theta_{\tilde{g}}}}
\newc{\as}{\alpha_s}

\title{A Light Bottom Squark in the MSSM}

\author{Athanasios Dedes and Herbi K. Dreiner\\ 
Rutherford Appleton Laboratory\\ 
Chilton, Didcot, OX11 0QX, UK\\ 
E-mail: \email{a.dedes@rl.ac.uk}, \email{h.k.dreiner@rl.ac.uk}}

\abstract{We study the compatibility of a light bottom squark
  $M_{{\tilde b}}<{\cal O}(10\gev)$ in the unconstrained MSSM. We
  consider the one-loop radiative corrections which are large for a
  heavy gluino ($>\!\!{\cal O}(150\gev)$). We then consider the
  renormalization group flow up to the Grand Unified scale. For most
  regions of the parameter space with a light sbottom we find colour
  and charge breaking minima. Only a small window in gluino mass and
  $\tan\beta$ is consistent with all bounds. This is alleviated by a
  light gluino, which is however only marginally experimentally
  allowed.}
 
\keywords{Beyond Standard Model, Supersymmetric Standard Model}

\preprint{hep-ph/0009001}

\begin{document}

\section{Introduction}

The experimental bound from LEP on the lightest supersymmetric
particle (LSP), assuming it is a neutralino, is given by
\cite{opallsp} 
\beq 
M_{{\tilde\chi}^0_1}> 40.9 \gev; \qquad ({\rm OPAL}),
\eeq 
and similar numbers from DELPHI ($36.7\gev$) \cite{delphilsp}, L3
($38.2\gev$) \cite{l3lsp} and ALEPH ($37.5\gev$) \cite{alephlsp}. This
bound assumes the supersymmetric grand unified relation between the
gaugino masses at the weak scale: $M_1=(5/3)\tan^2\theta_W\, M_2$
(where $\theta_W$ is the electroweak mixing angle) and employs the
chargino search. In a recent paper \cite{karmen} it was shown that if
you drop this theoretical assumption a LSP neutralino even as light as
$34\,MeV$ is consistent with all experiments.\footnote{For a
discussion of astrophysical bounds see \cite{karmen,supernovae}.}

It is the purpose of this letter to do a similar study for light
bottom squarks. Light top squarks have been extensively studied
elsewhere \cite{stop}. For large values of $\tan\beta$ (the ratio of
the vacuum expectation values of the two neutral CP-even Higgs bosons
in the MSSM) it can be natural to have light bottom squarks as well,
as we discuss in more detail below. Both the D0 and CDF experiments
have performed direct searches for the lightest bottom squark
\cite{d0,cdf} obtaining
\barr
m_{{\tilde b}}&>& 115\gev,\quad ({\rm D0})\\
m_{{\tilde b}}&>& 146\gev,\quad ({\rm CDF}).
\earr 
We have given the maximum bound which is obtained for a vanishing
neutralino LSP mass. In general the bound depends on the LSP mass and
becomes less sensitive as the mass difference between the LSP and the
squark is decreased. For smaller mass differences and also for smaller
bottom squark masses the LEP searches \cite{opalsbottom,delphisbottom1, 
delphisbottom2,l3sbottom,alephlsp} are more sensitive. However, even
in this case there remains a gap at very small mass differences to the
LSP which becomes more pronounced for very small squark masses
$m_{{\tilde b}}<{\cal O}(10\gev)$ \cite{lowmass}. At such low masses
the decay ${\tilde b}\ra b{\tilde \chi }^0_1$ is kinematically
suppressed by the final state quark even for vanishing neutralino
mass. A dedicated search for the top squark with a small mass
difference ($\Delta M$) to the LSP has been performed
\cite{smalldelta} reaching as low as $\Delta M=1.6\gev$; the threshold
for the decay ${\tilde t}\ra {\tilde\chi}^0_1c$. We are not aware of
such a search for light bottom squarks.

Light squarks can directly contribute to the hadronic cross section at
the $Z^0$ peak. As we will discuss below, the doublet and singlet
squarks mix and for specific mixing parameters the coupling to the
$Z^0$ can even vanish. Thus this constraint restricts the range of
sbottom mixing but can not exclude a light sbottom. This constraint
turns out to be very mild since the light sbottom is dominantly
an $SU(2)_L$ singlet.

Very light bottom squarks have recently been investigated in Refs.
\cite{pacetti,weiglein}. In \cite{pacetti} a possible influence on the
parameter $R(s)=\sigma(e^+e^-\!\ra {\rm hadrons})/\sigma(e^+e^-\!\ra
\mu^ +\mu^-)$ was studied. For a $b$-squark the asymptotic
contribution is only 1/12. This is 1/4 that of a b-quark due to the
missing spin degeneracy and is below the experimental sensitivity
\cite{pdg}. In \cite{weiglein} the effect of a light b-squark on the
electroweak precision data and on the MSSM Higgs sector was
investigated. It was found to be consistent with the precision data
provided the scalar top quark is not too heavy. The upper bound on the
lightest CP-even Higgs boson in the MSSM is slightly lowered.

In the following we discuss the theoretical implications of a very
light bottom squark. We focus on the embedding into the MSSM. We first
study the pole mass of the bottom squark at one-loop, including in
particular radiative corrections from the gluino which are large, and
also corrections from top and bottom Yukawa couplings to the Higgs
boson masses. We then study the renormalization group flow of the
right-handed bottom squark mass squared for both universal and
non-universal scalar fields at the GUT scale. We consider the
constraints from colour and charge breaking minima (CCB). The
constraints are relaxed by a light gluino. We finish with a brief
discussion of bounds on a light gluino, before we conclude.

\section{Parameters and Constraints}

In the MSSM there are two bottom squarks. The $SU(2)_L$ current
eigenstates are denoted ${\tilde b}_L,\;{\tilde b}_R$, where ${\tilde
b}_L$ is a doublet squark and ${\tilde b}_R$ is a singlet squark.
The corresponding states for the top squark are ${\tilde t}_L,\;
{\tilde t}_R$. The mass matrix of these squarks in the current
eigenstate basis is given for example in \cite{phases} and 
the one-loop radiative corrections are given in \cite{pierce}.

The mass eigenstates depend on the following parameters in the
standard MSSM notation \cite{review}: $M^2_{{\tilde Q},{\tilde
D},{\tilde U}}$, the doublet and singlet soft-breaking squark masses,
respectively; $M_{W,Z}$, the gauge boson masses; $\tan\beta$, the
ratio of the vacuum expectation values of the two neutral CP-even
Higgs fields; $A_b,\;A_t$, the tri-linear soft breaking terms; $\mu$,
the Higgs mixing parameter; and $M_{\tilde g}$, the gluino mass.  In
the following $m_b$ and $m_t$ denote the bottom and top quark mass,
respectively. We shall denote the lighter bottom squark ${\tilde b}_2$
and the heavier one ${\tilde b}_1$ in accordance with \cite{pierce}.
The scalar bottom mixing angle we denote $\theta_{\tilde b}$. All the
above parameters are considered to be ${\overline {DR}}$-running
parameters.

Besides the direct searches we have discussed in the introduction a
light b-squark would also contribute to the hadronic cross section at
the $Z^0$ peak, $\sigma_{had}^0$. The experimental bound for any
contribution beyond the SM is \cite{pdg,lepewwg} \footnote{We note 
that the Standard Model prediction of $\sigma_ {had}^0$ is currently
$1.7\,\sigma$ below the measured value. In principle this could be
exactly compensated by the light bottom squark \cite{weiglein}. In the
following we choose to focus only on the experimental {\it upper
bound} on a new contribution.}
\beq 
\Delta\sigma_{had}^0(Z^0)< 0.142\, nb, \quad(2\sigma).
\eeq 
%%%%%%%%%%%%%%%%%%%%%%%%%%%%%%%
At tree-level this requires the sbottom mixing angle to lie in the 
range (for $\sin\theta_W^2=0.2315$ and $N_c=3$)
\beq 
|\sin\theta_{\tilde b}|<0.535.  
\lab{lepbound} 
\eeq
At $2\sigma$, zero mixing is consistent with the data. In our analysis
we have included the one-loop contribution to $\sigma_{had}^0$ from
the scalar bottom. We have only plotted points which are consistent
with the bound \eq{lepbound}. It turns out that this constraint has no
effect on Figs.~\ref{fig1}-\ref{fig3}.  For ${\tilde b}_2$ satisfying
\eq{lepbound} the heavier bottom squark, ${\tilde b} _1$, couples
unsuppressed to both the photon and the $Z^0$. In order to avoid
experimental bounds from LEP1 and LEP2 we must therefore require
\beq
m_{{\tilde b}_1}\gsim 200\gev.
\eeq
In our scans below, we shall employ this bound, as well.

A light sbottom contributes to the running of the strong coupling
$\alpha_s$ between $m_{\tau}$ and $M_{Z^0}$. In order to see whether
this is consistent with the data one must include the light sbottom
both in the determination of $\alpha_s$ in a given experiment and also
in the beta function. This is beyond the scope of this letter. However, 
it has been performed for a light gluino \cite{alphas,csikor}. The most 
recent study \cite{csikor} with the smallest experimental error in
$\alpha_s$ was able to exclude a light gluino at the $70\%\,C.L.$. The
contribution to the beta function at one-loop of a singlet sbottom is
$1/12$ that of a gluino. We thus expect the effect to be significantly
smaller and beyond present experimental sensitivity.

\section{Sbottom Pole Mass}

We now investigate the effect of radiative corrections on the sbottom
pole mass. $m_{{\tilde b}_2}$ depends at tree-level on the parameters
$M_{\tilde Q}$, $M_ {\tilde D}$, $A_b$, $\mu$ and $\tan\beta$. At
one-loop \cite{pierce}, there is a further dependence on the stop mass
parameters, $M_{\tilde g}$, and $M_A$\footnote{We do not include
chargino-neutralino corrections since they are small \cite{pierce}.
Also, the light Higgs mass has been set to $M_Z$. Variation of the
Higgs mass to its upper limit affects very weakly the light sbottom
mass.}  The dependence on the stop sector and the Higgs sector
parameters is weak and we fix them to $A_t=300\gev$, $M_{{\tilde t}_R}
=300\gev$ and $M_A=400\gev$.  We also fix the following SM parameters
at the $Z^0$ scale to: $m_t (pole)=175\gev,\,\sin^2\theta _w=0.2315,\,
m_b=2.9\gev$, and $\alpha_s= 0.12$. As we discuss now, the dependence
on $M_{{\tilde b}_R}$, $A_b$ and on $M_{\tilde g}$ is strong.

%%%%%%%%%%%%%%%%%% FIG1  %%%%%%%%%%%%%%%%%%%%%%%%%%%%
\FIGURE[t]{\centerline{\hbox{\epsfig{figure=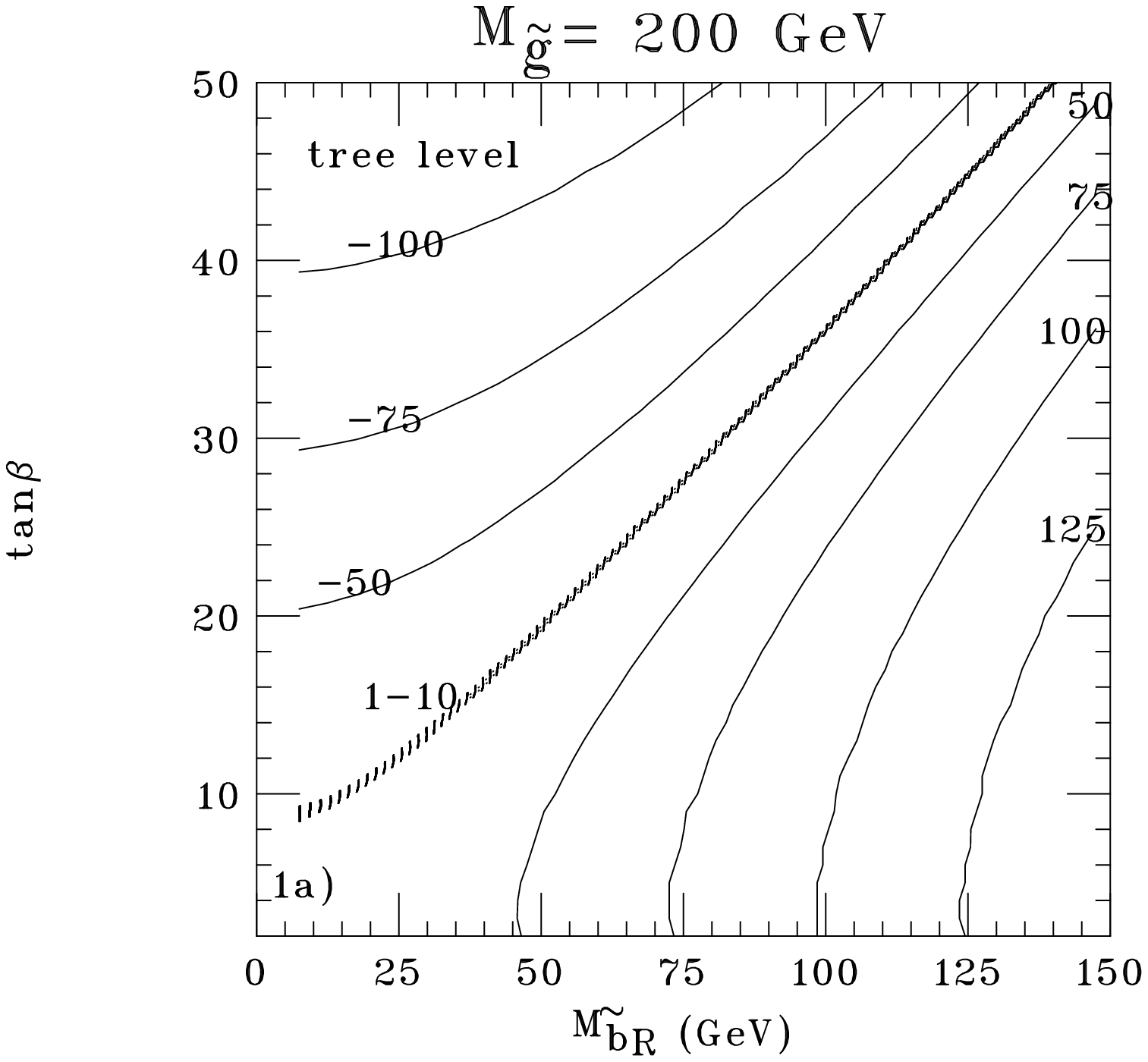,
height=2.9in,angle=90}
\epsfig{figure=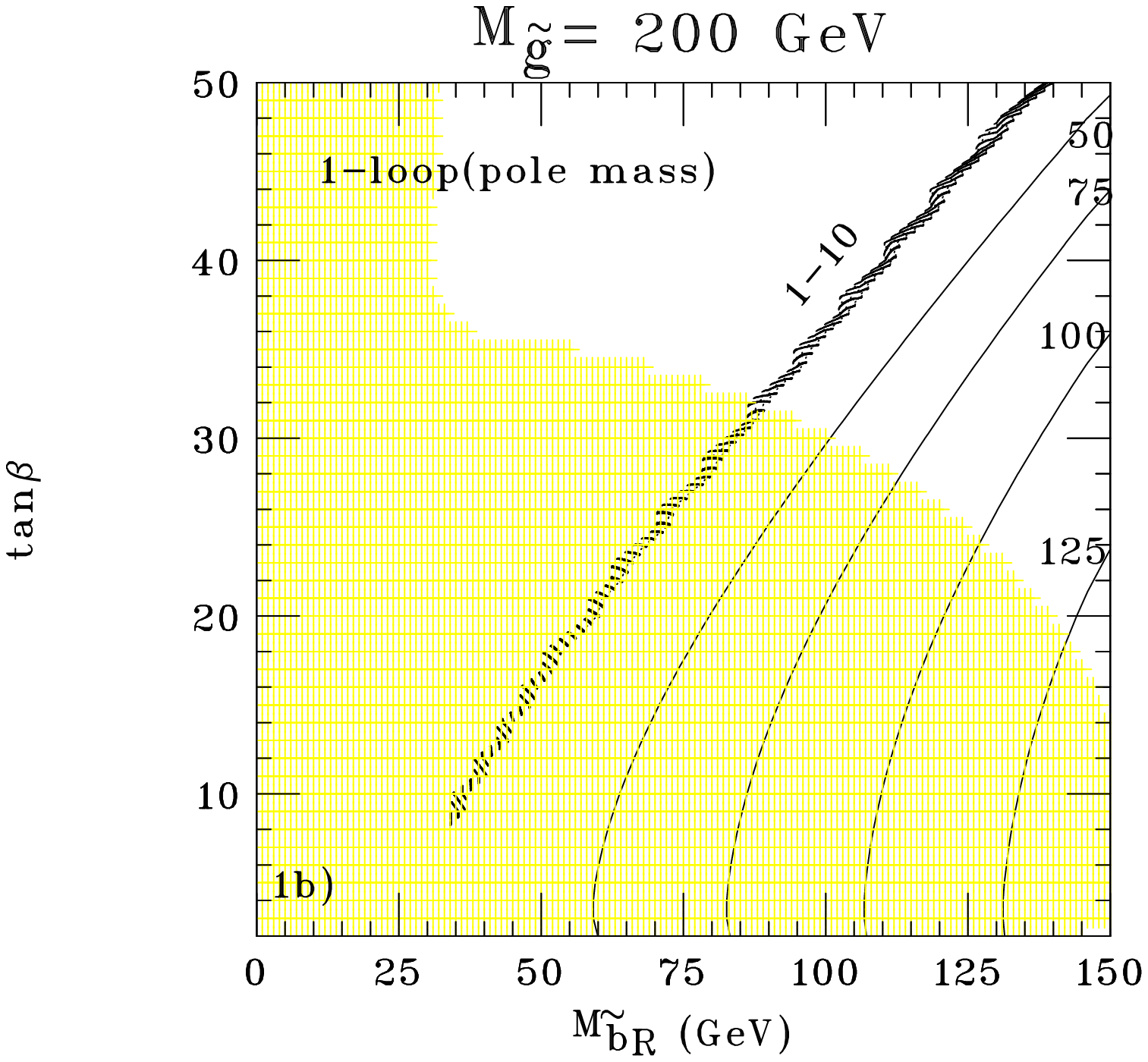,height=2.9in,angle=90}}}
\caption{ {\bf a)} Contour plot of the tree level light sbottom mass
$m_{\tilde{b}_2}$ as a function of the right handed singlet soft SUSY
breaking mass $M_{{\tilde b}_R}$ and $\tan\beta$.  The tree level
mass-squared $sign(m_{\tilde{b}_2}^2)|m_{\tilde{b}_2}^2| ^{1/2}$ is
getting negative for $\tan\beta \ge 10$ and for various values of
$M_{\tilde{b}_R}$ indicated in the figure.  The small dashed band
indicates values of the light sbottom in the region $1-10\gev$.  The
soft trilinear coupling $A_b$ has been set to zero in this plot. {\bf
b)} As in Fig.1a but for the physical 1-loop mass. Sbottom masses of
$50,\,75,\,100,\,125\gev$ are indicated for comparison. In the large
shaded region, the running $\overline{DR}$ mass squared ($M^2_{\tilde
{b}_R}$) is getting negative at a scale which lies below the GUT 
scale.}\label{fig1}}
%%%%%%%%%%%%%%%%%%%%%%%%%%%%%%%%%%%%%%%%%%%%%%%%%%%%%%%

%%%%%%%%%%%%%%%%%% FIG 2 %%%%%%%%%%%%%%%%%%%%%%%%%%%%
\FIGURE[t]{\centerline{\hbox{\epsfig{figure=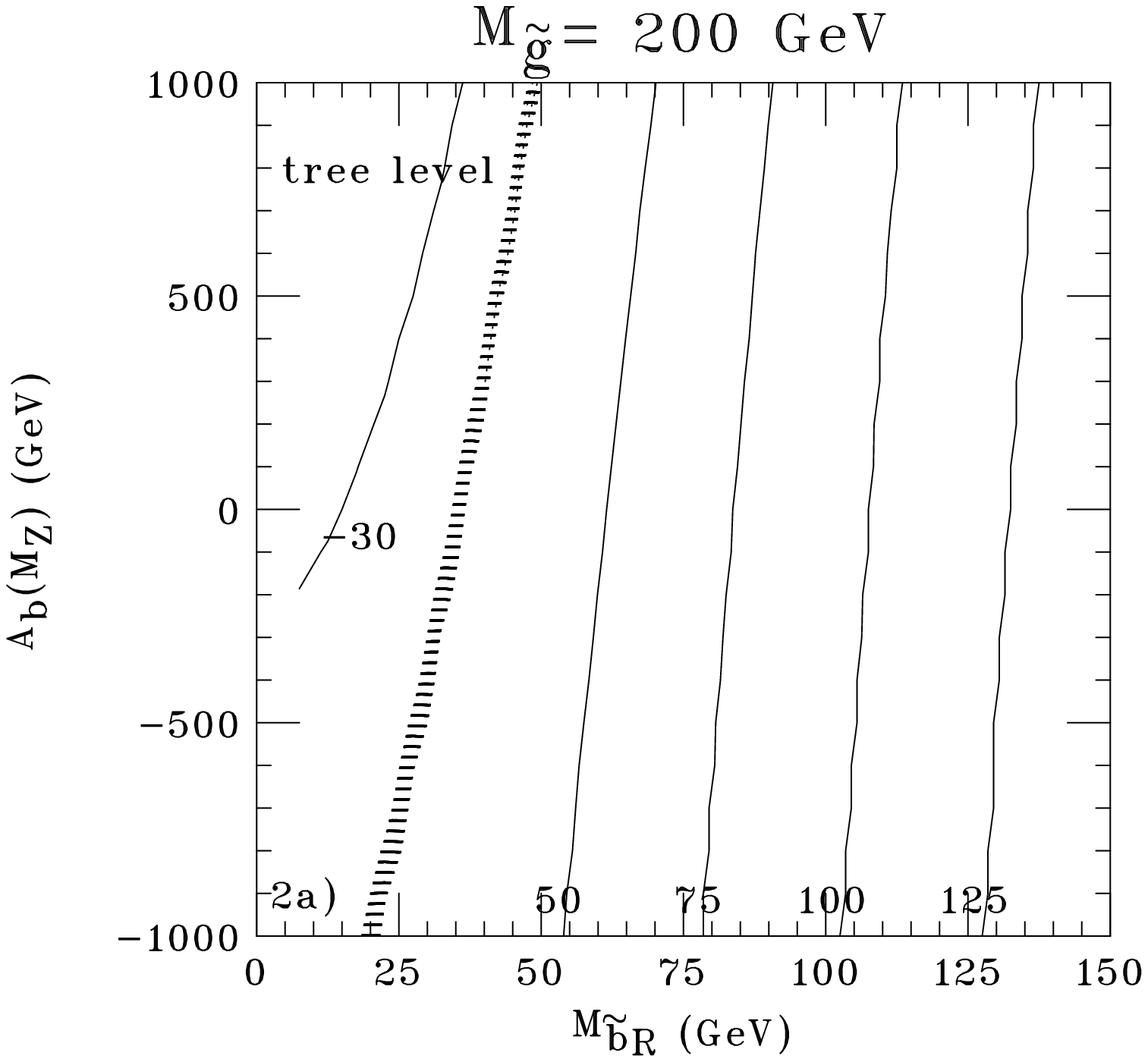,
height=2.9in,angle=90}
\epsfig{figure=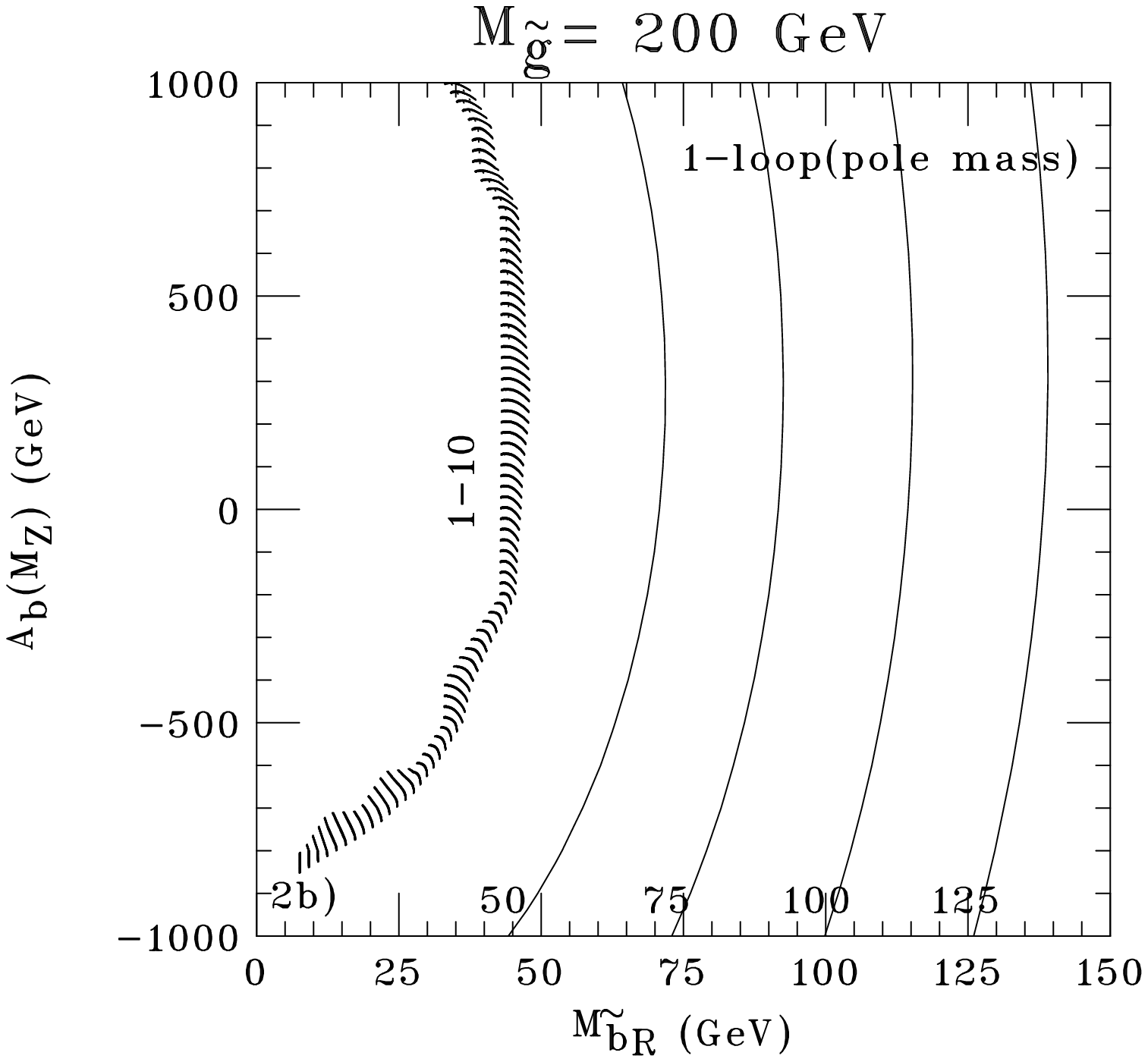,height=2.9in,angle=90}}}
\caption{ {\bf a)} Contour plot of the tree level light sbottom mass
as a function of the right handed singlet soft SUSY breaking mass
$M_{{\tilde b}_R}$ and the trilinear coupling $A_b(M_Z)$.  The value
of $\tan\beta$ is fixed to 15. {\bf 2b)} The same for the physical
1-loop light sbottom pole mass. The light sbottom mass contours, $1-10
\gev$ of this figure is completely within the shaded region of
Fig.1b. }\label{fig2}}
%%%%%%%%%%%%%%%%%%%%%%%%%%%%%%%%%%%%%%%%%%%%%%%%%%%%%%%

The first case we examine is that of a heavy gluino of $200\gev$, just
above the current experimental bound of $180\gev$ \cite{gluinobound}.
For now, we fix the remaining input parameters: $M_{\tilde{b}_L}=250
\gev, A_b=0\gev, \mu=250\gev$, at the $Z^0$ scale. In
Fig.~\ref{fig1}a,b we present contour plots of the lightest bottom
squark mass, $m_{\tilde {b} _2}$ in the ($M_{\tilde{b}_R
}$-$\tan\beta$) plane. We display both tree level and physical 1-loop
pole masses. The narrow shaded strip corresponds to masses in the
range $1-10\gev$. The area to the left of this narrow strip in Fig.~1b
is excluded since the scalar bottom pole mass squared turns out to be
negative. The effect of the radiative corrections is significant for
$\tan\beta\lsim15$. One can see up to a $40\gev$ difference between
the tree level and the 1-loop physical mass. They tend to push a fixed
sbottom mass to larger values of $M_{\tilde{b}_ R}$. The solution
region for a light sbottom is quite narrow and somewhat fine-tuned. A
variation of $1\gev$ of $M_{\tilde{b}_R}$ results in a variation of
more than $5\gev$ in the light sbottom mass in the ${\cal O}(<10
\gev)$ region.  Thus when determining the supersymmetric parameters
for a light bottom squark the radiative corrections need to be taken
into account.

With the above input values we obtain for the other (physical 1-loop)
masses : $m_{\tilde{b}_1}=255-300\gev$, $m_{\tilde{t}_1}=191-228\gev$,
$m_{\tilde{t}_2}=394-416 \gev$. All these masses satisfy the
current experimental bounds.

As we have already mentioned above, another parameter which plays a
crucial role in determining the mass of the bottom squark is the
trilinear coupling $A_b(M_Z)$. This parameter enters in both tree
level and 1-loop sbottom mass corrections and the effect of its
variation is presented in Fig.\ref{fig2}. The input parameter
$\tan\beta$ is fixed to $\tan\beta=15$. As we can see, the one loop
radiative corrections shift the mass contours by at most tens of
$\gev$. Large values of $A_b(M_Z)$ typically give smaller sbottom
masses.  The other physical masses ($m_{\tilde{b}_1},\,m_{\tilde{t}
_1},\, m_{\tilde{t}_2}$) vary inside the region we mentioned in the
previous paragraph.

%%%%%%%%%%%%%%%%%% FIG3  %%%%%%%%%%%%%%%%%%%%%%%%%%%%
\FIGURE[t]{\centerline{\hbox{\epsfig{figure=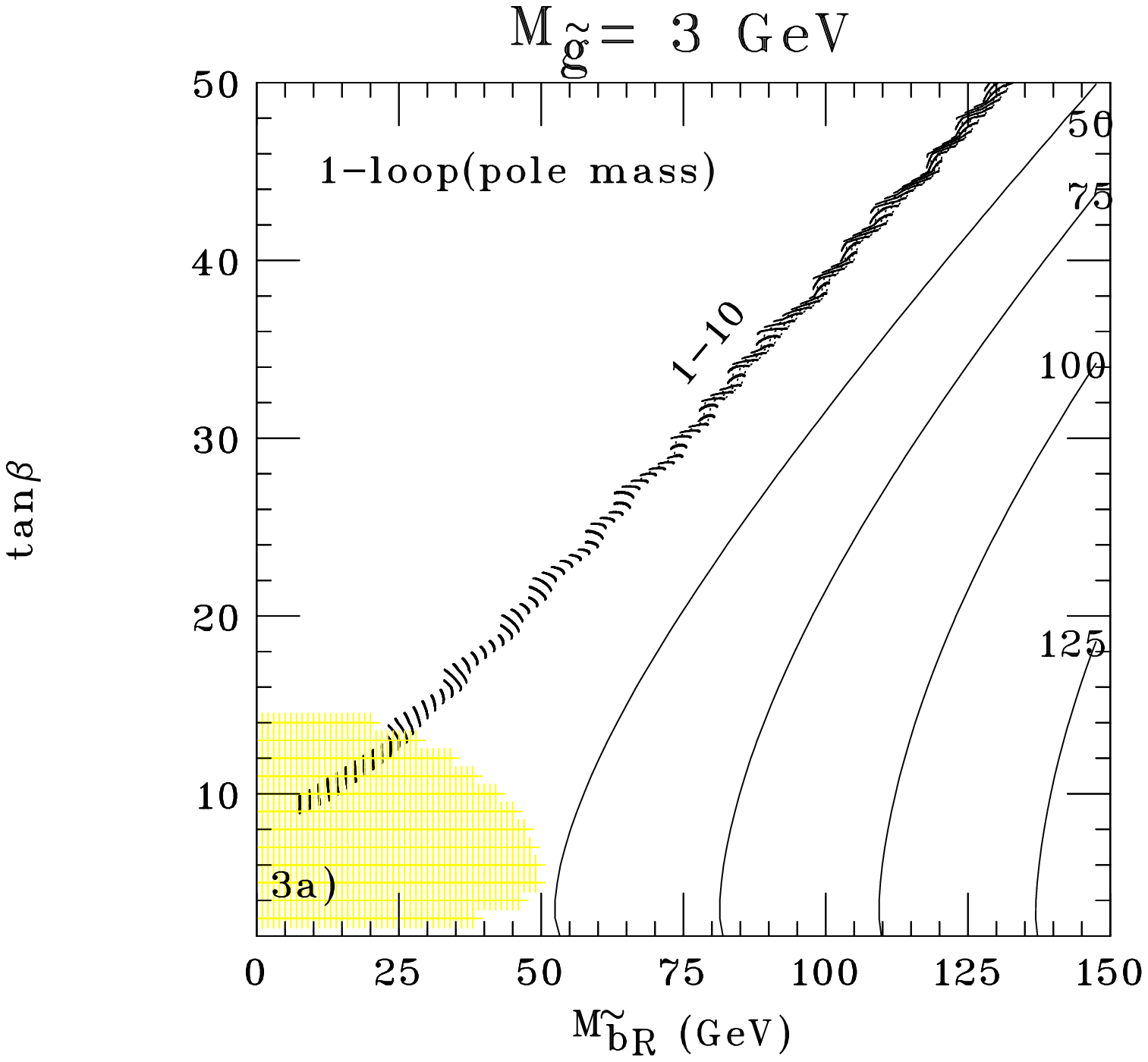,
height=2.9in,angle=90}, 
\epsfig{figure=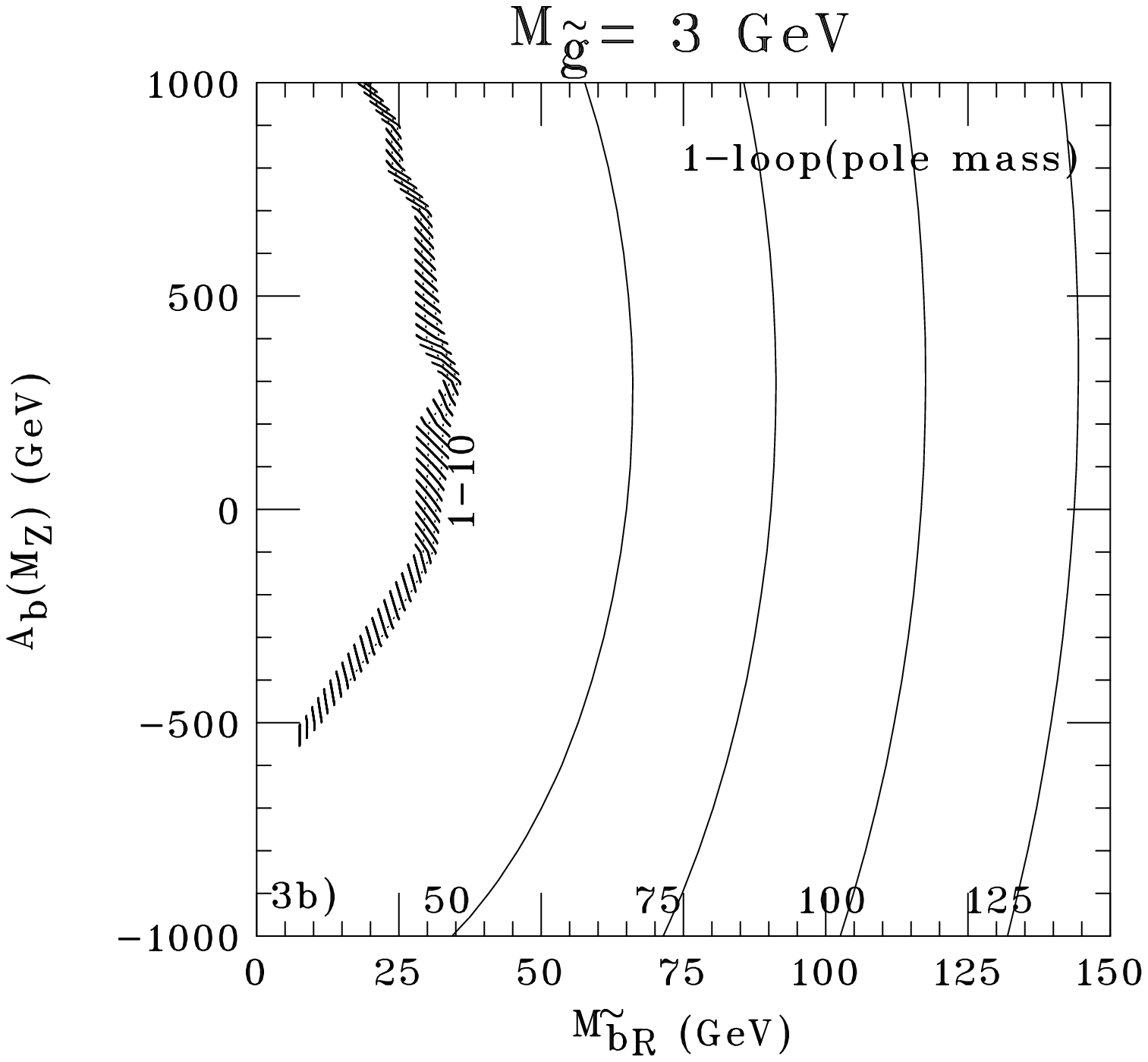,height=2.9in,angle=90}}}
\caption{ {\bf a)} The same as in Fig.1b  {\bf 3b)} and in Fig.2b
but for  a light gluino of mass, $M_{\tilde{g}}=3$ GeV.
The tree level results are those given in Fig.1a and Fig.2a .}
\label{fig3}}
%%%%%%%%%%%%%%%%%%%%%%%%%%%%%%%%%%%%%%%%%%%%%%%%%%%%%%%%%%

Further important radiative correction to the bottom squark masses are
those arising from the loops involving gluinos \cite{pierce}. This is
obvious when one compares the Figs.~\ref{fig1},\,\ref{fig2} where the
gluino mass is taken to be $200\gev$ with Fig.~\ref{fig3} where its
mass is set to be $3\gev$. We see that in the $(M_{{\tilde b}_R}$-$
\tan\beta)$ plane there is almost no effect from radiative corrections
for a light gluino compared to the tree-level result presented in
Fig.\ref{fig1}a. In the $(M_{{\tilde b}_R}$-$A_b)$ plane the effect is
much less dramatic for a light gluino, although there is still a
qualitative difference in the $A_b$ dependence compared to the
tree-level result.

In summary, a light bottom squark can be consistently implemented at
one-loop in the MSSM. The effect of radiative corrections as a function
of $M_{\tilde g}$ and $A_b$ is substantial, up to several tens of $\gev$
and must be considered when determining the supersymmetric parameters.

\section{Renormalization Group}
\subsection{Universal Scalar Masses at $M_X$}
We next consider the embedding of the MSSM in a more unified theory at
a high scale, $M_X={\cal O}(10^{16}\gev)$.  Having extracted the
$\ol{DR}$ quantity $M_{\tilde{b}_R}$ from the Figs~\ref{fig1},\,
\ref{fig2},\,\ref{fig3} we would thus like to see if these values are
compatible with the renormalization group running of the MSSM up to
$M_X$. In order to qualitatively understand the evolution we first
discuss an approximate analytic solution for the $M_{\tilde{b}_R}^2$
running mass. We present the full numerical analysis below.  The
renormalization group evolution of the $M_{\tilde{b}_R}^2$ mass is
given~\cite{rges} by,
%%%%%%%%%%%%%%%%%%%%%
\begin{eqnarray}
16 \pi^2 \frac{d M_{\tilde{b}_R}^2}{dt}=4 Y_b^2 \Sigma_b^2-
\frac{32}{3}g_3^2 M_3^2 -\frac{8}{15}g_1^2M_1^2 +
\frac{2}{5}g_1^2 {\rm Tr}(Y m^2) \;,
\label{eq4.1}
\end{eqnarray}
%%%%%%%%%%%%%%%%%%%
where 
%%%%%%%%%%%%%%%%
\begin{eqnarray}
\Sigma_b^2&\equiv& M_{H_d}^2+M_{\tilde{b}_L}^2+M_{\tilde{b}_R}^2
+A_b^2 \;,\\
{\rm Tr}(Y m^2)&\equiv& M_{H_u}^2-M_{H_d}^2 +\sum_{i=1}^{n_f}
(M_{\tilde{Q}_{L_i}}^2-2 M_{\tilde{u}_{R_i}^2} +
M_{\tilde{d}_{R_i}^2}-M_{\tilde{L}_{L_i}}^2+
M_{\tilde{e}_{R_i}^2}) \;.
\label{eq4.3}
\end{eqnarray}
%%%%%%%%%%%%%%%%%%%%
To start, we assume that the contribution from the bottom Yukawa
coupling is small, \ie we restrict ourselves to the region $\tan\beta
\lsim 10$. We also assume all of the squark, slepton and Higgs-boson
masses are the same at the GUT scale, \ie ${\rm Tr}(Y m^2)$ remains
zero at all scales.  In the case where the gluino is heavy,
$M_{\tilde{g}}= 200\gev$, the running of $M_{\tilde{b}_R}$ ``freezes''
below $M_{\tilde{g}}$ \cite{rges} to
%%%%%%%%%%%%%%%%%%%%%
\begin{eqnarray}
M_{\tilde{b}_R}^2= m_0^2 + C_3 + \frac{1}{9}C_1 -\frac{1}{3}
\sin^2\theta_w M_Z^2 \cos(2\beta) \;,
\label{rgerun}
\end{eqnarray}
%%%%%%%%%%%%%%%%%%
where $m_0$ is the common squark, slepton and Higgs-boson mass at the
GUT scale and
%%%%%%%%%%%%%%%%%%
\begin{eqnarray}
C_1(\mu)&=&-\frac{2}{11} M_1^2 \biggl [ 1-
\frac{\alpha_1^2(M_X)}{\alpha_1^2(\mu)} \biggr ] \;, \nonumber \\
C_3(\mu)&=&\frac{8}{9} M_3^2 \biggl [ 1-
\frac{\alpha_3^2(M_X)}{\alpha_3^2(\mu)} \biggr ] \;.
\end{eqnarray}
%%%%%%%%%%%%%%%%%
The D-term contribution, $-\frac{1}{3} \sin^2\theta_w M_Z^2
\cos(2\beta)$ in Eq.(\ref{rgerun}), is positive since $\cos(2\beta)$ is
negative, and also the bino contribution $C_1$ is positive. The
dominant term in the evolution of $M_{{\tilde b}_R}^2$ is the gluino
contribution, $C_3$.  Now, suppose that all the scalar masses at the
GUT scale are set to zero, $m_0=0$, and neglecting all the other
positive but small contributions (proportional to $\alpha_1^2$) we can
estimate the minimum mass of $M_{\tilde{b}_R}$ as
%%%%%%%%%%%%%%%%%%%%%
\begin{eqnarray}
M_{\tilde{b}_R}^2\gsim \frac{8}{9} M_3^2(M_3) \biggl [1-
\frac{\alpha_3^2(M_X)}{\alpha_3^2(M_3)} \biggr ].
\end{eqnarray}
%%%%%%%%%%%%%%%%%
 From the pole gluino mass, $M_{\tilde{g}}$,  we
extract the $\overline{DR}$ mass \cite{pierce},
%%%%%%%%%%%%%%%%%%%%
\begin{eqnarray}
M_3(M_3)=M_{\tilde{g}} \biggr [1-\frac{15 \alpha_3(M_3)}{4\pi} 
\biggr ] \;.
\end{eqnarray}
%%%%%%%%%%%%%%%%%
For $M_{\tilde{g}}=200\gev$, we get $M_3(M_3)=174\gev$  and
%%%%%%%%%%%%%%%
\begin{eqnarray}
M_{\tilde{b}_R} \gsim 150\gev.
\end{eqnarray}
%%%%%%%%%%%%%%%
For larger gluino masses this becomes larger.  If we allow for a
positive contribution from $m_0^2$ then $M_{\tilde {b}_R}$ becomes
correspondingly larger. From Figs.~\ref{fig1}, \ref{fig2} we see that
this value is incompatible with a light sbottom of ${\cal O}(<10\gev)$
in the small $\tan\beta\lsim 10$ region where the above solution is
valid. In fact, universality of the squark and slepton masses is
incompatible with all the values of $\tan\beta$. This is because
$m_0=0$ implies a small stop mass (even smaller than the sbottom one)
excluded by the current experimental data.

Let us now consider the case of a light gluino here taken to be 3 GeV.
In order to obtain chargino and neutralino masses compatible with the
experimental data we keep the common electroweak gaugino mass
$M_2=M_1=M_{1/2}$ at the GUT scale greater than 120 GeV. Then for
$M_{\tilde{g}}=3\gev$ and $\alpha_s(m_b)=0.22$ we get $M_3(M_3\simeq
m_b)=2.2\gev$. Evolving this up to the Z-scale using the relation
%%%%%%%%%%%%%%%%%%%
\begin{equation}
\frac{M_3(m_b)}{\alpha_3(m_b)}= \frac{M_3(M_Z)}{\alpha_3(M_Z)},
\end{equation}
%%%%%%%%%%%%%%%%%%%%%
we obtain $M_3(M_Z)=1.2\gev$  which in turn gives from (\ref{rgerun})
%%%%%%%%%%%%%%%%%%%
\begin{equation}
M_{\tilde{b}_R} \simeq m_0.
\end{equation}
%%%%%%%%%%%%%%%%%%%
That is compatible with the $M_{\tilde{b}_R}$ mass of our
Fig.~\ref{fig3} for positive $m_0$ at the GUT scale but again is not
compatible with the experimental bound on the top squark mass ($>120
\gev$). Thus we conclude here that a light sbottom mass of order
$\lsim{\cal O} (10\gev)$ is incompatible within the MSSM under the
assumption of universality of scalar masses as well as universality of
the electroweak gaugino masses at the GUT scale. The gaugino mass
universality is not essential.

\subsection{Non-Universal Scalar Masses at $M_X$}
Analytical solutions of the renormalization group equation
(\ref{eq4.1}) in the case of non-universal boundary conditions have
been obtained in~\cite{Nath}, under the assumption of a small bottom
Yukawa coupling and thus small $\tan\beta$ values. Even in this
approximation the results are quite complicated. The term ${\rm Tr}(Y
m^2)$ of (\ref{eq4.3}) is now non-zero at the GUT scale and below. One
must thus consider the running of the other soft masses as well. The
coupled system of differential equations is difficult to solve.

Since we are interested in solutions of the RGE's even in the large
$\tan\beta$-regime we solve them numerically. Instead of solving the
RGE's assuming a specific pattern for the soft breaking masses at the
GUT scale, we use our results from Figures~1b and 3a for the
$\overline{DR}$ right handed soft bottom mass at the Z-scale and run
this up to the GUT scale together with all the other masses and
couplings. We use two-loop RGE's for all the couplings and masses and
full treatment of threshold effects~\cite{dedes1}. All the other
parameters have been taken to satisfy the current experimental
constraints. 

The dominant effect on the running of $M_{\tilde{b}_R}^2$ is the
gluino mass. As we run the RGE's up in scale, the gluino mass drives
$M_{\tilde{b}_R}^2$ to negative values. The scale where
$M_{\tilde{b}_R}^2$ becomes negative depends on $\tan\beta$ and on the
initial $M_{\tilde{b}_R}^2$. Note that the positive bottom Yukawa
coupling contribution compensates the negative ones from the gluino
in (\ref{eq4.1}) \footnote{Non-universality affects also the results
through the term ${\rm Tr}(Ym^2)$ in (\ref{eq4.1})}.

$M_{\tilde{b}_R}^2<0$ implies a charge and colour breaking minimum of
the scalar potential (CCB). Note that the CCB we obtain is {\it above}
a given scale, \ie at the weak scale or below charge and colour
symmetry are restored. Such a CCB at high scales (as opposed to
present scales \cite{abel}) is not observationally excluded to our
knowledge \cite{subir}, however it would substantially alter the
conventional cosmology at high temperatures. If colour and charge are
broken close to the electroweak scale we would expect there to be
experimentally observable effects. It is beyond the scope of this
letter to investigate this in detail. In the following we shall
consider charge and colour breaking to be excluded for scales
$Q<1\,TeV$.

In order to avoid CCB we must go to large initial values of
$M_{{\tilde b}_R}^2$. As we saw in our approximate analytical
solution, the only possibility then of getting a light bottom squark
(pole) mass is to go to large values of $\tan\beta$.  This is
confirmed by our full numerical calculations. In the following Table
we summarize the lower bound on $\tan\beta$ for a given gluino mass
for which there is no CCB at all scales below $Q<1\,TeV$.
%%%%%%%%%%%%SAKIS HERE%%%%%%%%%%%%%%%%%%%%%%%%%%%%%%%%%%%%
In fact there is also an upper bound on $\tan\beta$. For large values
of $\tan\beta$ the light sbottom squared pole mass becomes negative.
In other words the gluino contribution to the physical sbottom mass is
such that there is always a tachyon. The upper bound on $\tan\beta$ is
also summarized in the Table below.
%%%%%%%%%%%%%%%%%%%%%%%%%%%%%%%%%%%%%%%%%%%%%%%%%%%%%%%%%%%%%%%%%
\beq
\begin{tabular}{|c|c|c|c|c|}\hline
$M_{\tilde g}$ &$200\gev$& $250\gev$&$300\gev$&$350\gev$ \\ \hline
$\tan\beta>$ &19  &23 & 28 & 33 \\ \hline
$\tan\beta<$ &53  &48 & 40 & 34 \\ \hline
\end{tabular}
\lab{table}
\nonumber
\eeq
%%%%%%%%%%%%%%%%%%%%%%%%%%%%%%%%%%%%%%%%%%%%%%%%%%%%%%%%%%%%%%%%%
So for $M_{\tilde g}=200\gev$ we only get a light bottom squark
without CCB for $\tan\beta>19$ and we avoid a tachyonic sbottom for
$\tan\beta<53$. For $M_{\tilde g}>350\gev$ there is no physical light
sbottom mass in the MSSM.  One could rather expect this result : A
light sbottom of order of 10 GeV or lighter is potentially unstable
under the radiative corrections of particles with mass of a few
hundred GeV.
%%%%%%%%%%%%%%%%SAKIS HERE%%%%%%%%%%%%%%%%%%%%%%%%%%%%%%%

Next, we consider the case where we require that there is no CCB for
all values $Q<M_X$, \ie up to the GUT scale. The corresponding
excluded parameter range is shown as the large shaded region in
Figs.~1b and 3a for a heavy and light gluino, respectively. For a
heavy gluino, we see that in order to retain the conventional
cosmology we must require $\tan\beta>32$ and $M_{{\tilde b}_R}>90
\gev$. For a heavier gluino, $M_ {\tilde g}>300\gev$, the entire 
$\tan\beta <50$ plane is excluded. For a light gluino (Fig~3a) this
excluded region is substantially reduced. Also $M^2_{\tilde{b}_R}(Q)$
turns negative only very close to the GUT scale at about $Q\sim
5\times 10^{15}\gev$ thus possibly avoiding most cosmological
constraints.

In summary, in the case of a heavy gluino in order to have a light
bottom squark we typically obtain CCB. The heavier the gluino the
lower the scale at which CCB is obtained. CCB below $1\,TeV$ is only
avoided for large values of $\tan\beta$, as summarized in \eq{table}.
In order to avoid CCB altogether we must go to very high values of
$\tan\beta$ and $M_{{\tilde b}_R}$ as shown in Fig. 1b. These
constraints are largely avoided for a light gluino, as seen in
Fig. 3a.

So far we have not considered radiative electroweak symmetry breaking
(RESB) \cite{ibanez}. This is potentially a very strict requirement.
As we saw, for a heavy gluino we required large $\tan\beta$ in order
to obtain a light bottom squark. This leads to a large bottom quark
Yukawa coupling such that possibly both $M^2_{H_u}$ and $M^2_{H_d}$
are negative. This is inconsistent with electroweak symmetry breaking.
A systematic check of this constraint is beyond the scope of this
letter. However, we would expect it to possibly be important.

%%%%%%%%%%%%%%%%%%%%%%%%%SAKIS ENDED HERE%%%%%%%%%%%%%%%%%

\section{Light Gluino}
As we have discussed, the introduction of a light gluino naturally
allows for a light $M_{{\tilde b}_R}$ even for small values of
$\tan\beta$, while avoiding any CCBs. However, it appears that such a
light gluino is experimentally excluded. In order to discuss this we
distinguish between a decaying and a non-decaying light gluino. The
latter could for example be the LSP. We first discuss the case of a
decaying light gluino.

Until fairly recently, there was a window in the search for a decaying
light gluino after combining several sets of experimental data
\cite{experiment,summary}. However, this window has now been closed
\cite{glennys} by new data from KTeV \cite{ktevgluino} and from LEP
\cite{lepgluino,csikor}. We do not consider it any further.

Next we consider a stable light gluino. (It would be stable if it were
the LSP.)  It would have been produced in the early universe and would
have a non-vanishing relic density today. These relic gluinos could
bind with nuclei (possibly after forming a bound state, such as
$R^0\equiv{\tilde g}g$) leading to anomalously heavy nuclei. The
number of such nuclei depends on the relic density (which is a
function of the self-annihilation cross section) and also on the
binding potential with nuclei (which depends on the scattering cross
section with nuclei). The relic density was first considered in
\cite{gluinorelic}. The resulting number of anomalously heavy nuclei
present today were shown to be excluded by existing searches in
\cite{ellisgluino}.  More recently this problem has been revisited
with more detailed work on the binding potential with nuclei
\cite{mohapatragluino}, however with the same conclusion, excluding a
stable gluino \cite{mohapatra}. 

In \cite{baer} the self-annihilation cross section of the gluinos was
reinvestigated. The authors concluded that unknown non-perturbative
effects could possibly lead to a larger cross section and thus a
significantly smaller relic density. This could possibly avoid the
bounds from anomalous heavy nuclei searches. The authors take this as
a motivation to re-examine bounds from colliders.  Using existing
analyses from LEP (OPAL \cite{opalgluino}) and the Tevatron (CDF
\cite{cdfgluino}) they exclude the full range of gluino masses from
$3\gev-130\gev$.\footnote{There is a possible window between $25\gev
-35\gev$ for an ``unlikely'' set of parameters \cite{baer}, which is
not of direct interest to our problem. The phenomenological
consequences of this window have been further explored in \cite{raby}.}

\section{Conclusions}

We have investigated the case of whether a light bottom squark ${\cal
O}(<10)$ GeV can be accommodated in the unconstrained MSSM. In our
analysis we have included all the relevant one-loop corrections to the
physical sbottom pole mass. For $\tan\beta\lsim15$ these corrections
are large and need to be included when determining the physical
parameters. The main effect is from the heavy gluino mass, but also
the trilinear coupling $A_b$ leads to significant effects. In this
precise framework, we were able to extract the running parameters and
evolve them up to higher scales with the full two-loop RGEs including
all threshold effects. In detail we find:

\begin{itemize}

\item If we assume universal scalar masses at the GUT scale (minimal
supergravity scenario) we find a light sbottom is inconsistent with
the experimental bounds on the other supersymmetric scalars. Thus
in this scenario a light sbottom is excluded.

\item  A light sbottom can be embedded in the MSSM with non-universal 
scalar boundary conditions at the GUT scale only for specific
conditions. For a heavy gluino ($M_{\tilde g}>180$ GeV) it requires
large values of of $\tan\beta>30$, in order to avoid CCB at scales
below $Q<1\,TeV$. This lower bound on $\tan\beta$ grows with the
gluino mass, \eq{table}. For each gluino mass there is also an {\it
upper} bound on $\tan\beta$ beyond which the light sbottom mass
becomes tachyonic, \eq{table}. Above $M_{\tilde g}>350\gev$ a light
sbottom is completely excluded. The gluino mass is thus restricted to
the range $180\gev < M_{\tilde g}<350\gev$.

\item If we require the absence of CCBs up to the GUT scale the
allowed values of $M_{\tilde g}$ and $\tan\beta$ are significantly
more restricted as summarized in Fig. 1b. Gluino masses above
$300\gev$ are already excluded.

\item A light sbottom could be embedded naturally in the
MSSM with a light gluino $\sim 3 GeV$ in a less fine tuned way
avoiding also CCB constraints for almost all the $\tan\beta$ values.
However, a light gluino seems to be experimentally unlikely.

\end{itemize}

We conclude that a light sbottom hypothesis is not completely  excluded
in the MSSM but it is disfavoured. 

\vspace*{1cm}

{\it Note Added:} 
  After completing this paper, hep-ph/0008321 by Plehn and 
  Nierste was put on the net. It is complementary to our work 
  focusing on effects in the B-meson data for specific 
  bottom decays to sbottoms.

\vspace{0.8cm}

\noindent{\bf \large Acknowledgments}\vspace{0.4cm}

\noindent {\small We would like to thank G. Weiglein and Sven 
Heinemeyer for discussions on his related paper and R.G. Roberts for
discussions on the running of $\alpha_s$. We thank John Ellis, Glennys
Farrar, R. Mohapatra and Keith Olive for discussions of the light
gluino. We also thank the RAL experimentalists Bill Murray, Rob
Edgecock and Alberto Ribbon for discussions on the recent experimental
LEP and CDF/D0 searches. We both thank the Rutherford Laboratory for
an excellent working atmosphere and a great time. This work was
completed despite poor treatment by PPARC.}

\end{document}